\documentclass[eqsecnum,superscriptaddress,twocolumn,nobibnotes, aps, reprint, prd, nofootinbib]{revtex4-2}
\usepackage[usenames,dvipsnames]{color}
\usepackage{amsmath,amsthm,amsfonts,amssymb,amscd,amsbsy}
\usepackage[justification=justified,singlelinecheck=false]{caption}
\usepackage{subcaption}
\usepackage{bbm}
\usepackage{graphicx}
\usepackage{booktabs}
\usepackage{array}
\usepackage{xcolor}
\usepackage{url}
\usepackage[nodayofweek]{datetime}
\usepackage[english]{babel}
\usepackage{mathtools}
\usepackage[toc,page]{appendix}
\usepackage{verbatim} % *** Multi-line comments
\usepackage{amsthm} % *** Theorems & definitions
\usepackage{enumerate}
\usepackage[shortlabels]{enumitem}
\usepackage{bbm} %math symbols
\usepackage[normalem]{ulem} % *** Dashed & pointed underline
\usepackage{bm}
\usepackage{setspace}
\usepackage{hyperref}
\usepackage{cleveref}
\usepackage{amsmath}
\usepackage{extarrows}
\usepackage{fancyhdr}
\usepackage{simplewick}
\usepackage{marvosym}
\usepackage{slashed}
\usepackage{tikz}
\usepackage[compat=1.1.0]{tikz-feynman}

\DeclareMathOperator{\SU}{SU}
\DeclareMathOperator{\U}{U}
\DeclareMathOperator{\Sp}{Sp}
\DeclareMathOperator{\trace}{Tr}
\renewcommand{\Im}{\text{Im}}
\renewcommand{\Re}{\text{Re}}

\graphicspath{{.},{./Figures}}

\begin{document}

\title{Leptogenesis During an Era of Early SU(2) Confinement}

\author{India Bhalla-Ladd}
\email{ibhallal@uci.edu}
\affiliation{Department of Physics and Astronomy, University of California, Irvine, CA 92697, USA}
\affiliation{Department of Logic \& Philosophy of Science, University of California, Irvine, CA 92697, USA}
\author{Izzy Ginnett}
\email{iginnett@uci.edu}
\affiliation{Department of Physics and Astronomy, University of California, Irvine, CA 92697, USA}
\author{Tim M.P. Tait}
\email{ttait@uci.edu}
\affiliation{Department of Physics and Astronomy, University of California, Irvine, CA 92697, USA}
\date{\today}%

\begin{abstract}
    We explore leptogenesis during a cosmological epoch during which the electroweak $\SU(2)_L$ force is confined. During weak confinement, there is only one conserved non-anomalous global charge, $r$, which is a linear combination of lepton-number, baryon-number, and hypercharge.  The inclusion of heavy Majorana neutrinos leads to an $r$-charge and $CP$-violating interaction, allowing for the generation of an $r$-charge asymmetry, which translates into a baryon asymmetry post $\SU(2)_L$ deconfinement. Determining the resulting baryon asymmetry as a function of the model parameters, we find that the predicted baryon-asymmetry can match observations for a wide swath of parameter space. While leptogenesis under the assumption of a standard cosmology relies on the complex phase of the neutrino Yukawa couplings, the asymmetry generated in this novel background cosmology primarily depends on a strong phase from $\SU(2)_L$ confinement and favors negligible $CP$-violation in the right-handed neutrino decays.    
\end{abstract}

\maketitle
% UNCOMMENT BELOW IF WE WANT A TABLE OF CONTENTS
% \tableofcontents

%%%%%%%%%%%%%%%%%%%%%%%%%%%%%%%%%%%%%%%%%%%%%%%%%%%%%%
\section{Introduction}
\label{sec:introduction}
%%%%%%%%%%%%%%%%%%%%%%%%%%%%%%%%%%%%%%%%%%%%%%%%%%%%%%

The observed matter-antimatter asymmetry of the Universe remains an unsolved mystery tying together cosmology and particle physics.
Experimental measurements of the cosmic microwave background (CMB) and primordial light elements produced via big bang nucleosynthesis (BBN) agree with one another: 
\begin{equation}
\begin{split}
    Y_{\Delta B}^{\text{BBN}} & = (8.10 \pm 0.85) \times 10^{-11}, \\
    Y_{\Delta B}^{\text{CMB}} & = (8.79 \pm 0.44) \times 10^{-11},
\end{split}
\label{eq:YDB-experiment}
\end{equation}
where uncertainties are quoted at a 95\% confidence interval \cite{csfong2013}.  Famously, Sakharov determined that there are three conditions that must be satisfied in order to generate a baryon asymmetry from an initially symmetric Universe: (1) $B$-violation, (2) $C$ and $CP$-violation, and (3) a departure from thermal equilibrium \cite{Sakharov_1991}.  Although all of Sakharov's conditions can be present within the Standard Model (SM), the parameters of the SM do not allow it to reproduce the observed $Y_{\Delta B}$, and thus models of ``baryogenesis" require the addition of physics beyond the SM. 

A popular class of solutions realizes baryogenesis via leptogenesis, in which an asymmetry is initially produced in lepton number, and is converted into a baryon asymmetry via the electroweak sphalerons \cite{Fukugita1986}.  Typically, Majorana $\SU(2)_L$-singlet neutrinos $N_R$ are added to the SM, interacting with the active $\SU(2)_L$ doublet neutrinos through Yukawa interactions,
\begin{equation}
    %\mathcal{L}_I \supset 
    i\overline{N}_{R_\beta}\slashed{\partial}N_{R_\beta} - \frac{1}{2}M_{R \beta}\overline{N^c}_{R_\beta} N_{R_\beta} - \epsilon_{ab}\lambda_{\beta\alpha}\overline{N}_{R_\beta}\ell_\alpha^aH^b + \text{h.c.} 
    \label{eq:Lseesaw}
\end{equation}
where $\ell_\alpha^a = (\nu_\alpha, \alpha)$ is the $\SU(2)_L$-doublet lepton of flavor $\alpha$, $H^b = (H^+, H^0)$ is the $\SU(2)_L$-doublet Higgs, and $\beta$ labels the $N$ and flavor. The same ingredients amend the SM to allow for non-zero neutrino masses via the see-saw mechanism. The physical mass eigenstates are linear combinations of $N$ and $\ell$, which mix via the Dirac masses $m_D \equiv \lambda v$ induced when the Higgs acquires a vacuum expectation value $v$.
Assuming that $M_R$ is much greater than $m_D$, and being cavalier with flavor indices, the physical neutrino masses are:
\begin{equation}
\begin{split}
    m_2 & = \frac{M_{R}+\sqrt{M_R^2+4m_D^2}}{2} \approx M_R, \\
    m_1 & = \frac{M_{R}-\sqrt{M_R^2+4m_D^2}}{2} \approx \frac{m_D^2}{M_R} .
\end{split}
\end{equation}
Provided the parameters $M_R$ and $\lambda_{\beta \alpha}$ are chosen appropriately, $m_1$ can be made to match realistic neutrino masses.  In particular, the tiny experimentally observed mass splittings favor large $M_R$, which can allow for order one $\lambda_{\beta \alpha}$.  Both the $M_R$ and $\lambda$ are complex parameters, and for three generations generically predict $CP$-violation in the neutrino sector.  Provided that the $N_R$ decay out-of-equilibrium, all three Sakharov conditions are satisfied, and will produce a lepton asymmetry that is then converted into a baryon asymmetry by the electroweak sphalerons, successfully producing the observed baryon asymmetry of \cref{eq:YDB-experiment} for heavy masses. Constraints on the parameter space were explored in Ref.~\cite{Granelli_2025} by numerically solving the density matrix equations for the standard leptogenesis scenario, including flavor effects. Their analysis favors $M_R \sim 10^9$ GeV for fine-tuning $\Delta \lesssim 10$, well outside of the direct reach of current experiments. Many compelling variants of standard leptogenesis have been explored, such as Dirac leptogenesis, which works in the limit of $M_R\rightarrow 0$ \cite{Dick2000_physrev}, and relies on additional new physics (such as massive scalar particles \cite{Heeck2023_physrevd}) to store lepton number in the right-handed neutrinos.

The expectations for the parameters which would allow the see-saw theory of neutrino masses to successfully realize leptogenesis are strongly dependent on the extrapolation of the conditions in the early Universe from much more recent observations of its state.  As a result, there is a huge uncertainty in those expectations, and a novel cosmological history could retain consistency with observation while favoring dramatically altered see-saw parameters to explain leptogenesis.
We explore one such cosmology, the ``Weak Confined Standard Model" (WCSM), in which there is an early period where the weak force is stronger than it is today, strong enough that it confines the SM quark, lepton, and Higgs doublets into $\SU(2)$-neutral composite ``hadrons". This can be realized via a scalar field $\phi$ interacting with the $\SU(2)$ gauge fields:
\begin{equation}
    \mathcal{L} = -\frac{1}{2}\left( \frac{1}{g^2}-\frac{\phi}{M} \right)\trace[W_{\mu\nu}W^{\mu\nu}]-V(\phi) ,
    \label{eq:phiL}
\end{equation}
where $g$ is the $\SU(2)_L$ gauge coupling and $W^{\mu\nu}$ is the $\SU(2)_L$ field strength tensor. A nonzero vacuum expectation value (vev) for the scalar field induces an effective $\SU(2)$ coupling,
\begin{equation}
    \frac{1}{g^2_{\text{eff}}}=\frac{1}{g^2}-\frac{\langle \phi \rangle}{M} .
\end{equation}
For $\langle\phi\rangle \sim \frac{M}{g^2}$, $g_\text{eff} > g$ and the weak force becomes stronger. The potential for $\phi$ can be engineered such that $\langle\phi\rangle$ allows for a large $g_\text{eff}$ in the early universe, but transitions to a small value before electroweak symmetry breaking \cite{Ipek:2018lhm,Croon:2019ugf}. As a result, $\SU(2)_L$ can be confined at early times with a confinement scale denoted by $\Lambda_W$, whose value is controlled by the the dynamics of
the $\phi$ field in \cref{eq:phiL}.
In order to match on to cosmological measurements, $\SU(2)_L$ must deconfine later on when $\langle\phi\rangle$ shifts to a different value, at which point the cosmology becomes more standard.  Such a period of early $\SU(2)_L$ confinement has been shown to result in modifications of e.g. freeze-out of electroweakly-charged dark matter \cite{Howard:2021ohe}.

The effective field theory (EFT) describing the composite bound states and their interactions was analyzed in Ref.~\cite{Berger_2019}, based on the approximate $\SU(2N_f)$ chiral flavor symmetry, which lattice results \cite{Arthur2016_physrevd} suggest is spontaneously broken to $\Sp(2N_f)$.
This symmetry breaking pattern results in 65 pseudo-Nambu-Goldstone bosons (pNGBs) $\Pi$, whose tiny masses arise from the explicit breaking of the chiral symmetry by the SM gauge and Yukawa interactions, a spectrum of 12 composite fermions $\Psi$ with masses $\sim\Lambda_W$, and a scalar $\Phi$ whose mass is also $\sim\Lambda_W$.
During weak confinement, baryon- and lepton-number are not conserved quantities.  The only conserved non-anomalous global symmetry is $\U(1)_r$, defined as~\cite{Berger_2019}
\begin{equation}
    r \equiv -2Y+\sum_i\left( \frac{B}{3}-L_i \right) = -2Y+B-L .
    \label{eq:r-charge}
\end{equation}
As $r$ contains both baryon- and lepton-number, generating an asymmetry in it will ultimately lead to a baryon asymmetry.  However, this requires physics beyond the WCSM itself, which is provided by the right-handed neutrinos $N$.

We work with the standard see-saw theory of neutrino masses (with three singlet neutrinos), encapsulated by the Lagrangian of \cref{eq:Lseesaw}, and examine the requirements for successful leptogenesis if the Universe undergoes a period of early weak confinement.  The Majorana masses provide violation of $\U(1)_r$, which after deconfinement maps onto violation of lepton-number.  The Yukawa interactions $\lambda_{\beta\alpha}$ include a $CP$-violating phase, and the massive states resulting from the weak confinement can decay out-of-equilibrium.  We find that a sufficient baryon asymmetry can be generated, but for very different parameters than those necessary in a standard cosmology.

Our article is organized as follows. In \Cref{sec:su2-confined-theory}, we review the WCSM with the addition of three generations of Majorana $\SU(2)_L$-singlet neutrinos, focusing particularly on the ingredients salient for leptogenesis. In \Cref{sec:cosmology}, we discuss how this set-up generates a baryon-asymmetry, including a discussion of the Boltzmann equations governing the production of $Y_{\Delta B}$ in \Cref{sec:boltzmann-eq}, washout effects in \Cref{sec:washout}, and a computation of the rate of $CP$-violation in \Cref{sec:cp-asym}. Finally, we conclude and discuss further possible work in \Cref{sec:conclusion}.

%%%%%%%%%%%%%%%%%%%%%%%%%%%%%%%%%%%%%%%%%%%%%%%%%%%%%%
\section{EFT for SU(2) Confined Theory}
\label{sec:su2-confined-theory}
%%%%%%%%%%%%%%%%%%%%%%%%%%%%%%%%%%%%%%%%%%%%%%%%%%%%%%

We follow the notation of Berger et al.\cite{Berger_2019}, collecting the constituent particles of the WCSM into multiplets:
\begin{equation}
\begin{split}
    & \psi = \{ \ell_1,q_1^R,q_1^G,q_1^B,\ldots \},\\
    & \varphi = \{\epsilon H,-H^*\}, \\
    & \xi = \{ E_1,U_1^R,U_1^G,U_1^B,D_1^R,D_1^G,D_1^B,\ldots \}
\end{split}
\label{eq:fundamental-multiplers}
\end{equation}
where $\psi_i$ for $i=1,2,\ldots,4N_f$ contains the $\SU(2)_L$ doublet fermions, $\varphi_a$ for $a=1,2$ packages both the $\SU(2)_L$ Higgs doublet and its conjugate (for convenience), and $\xi=1,2,\ldots,7N_f$ contains the $\SU(2)_L$ singlet fermions. From here on, per the SM we take $N_f=3$ generations. 

\subsection{Vacuum and States}

During weak confinement, the spectrum of particles includes 65 composite pNGB scalars built out of two fermions ($\Pi_{ij}\sim \psi_i\epsilon\psi_j$), 12 Higgs-fermion composite fermions ($\Psi_{ia}\sim\psi_i\varphi_a$), and a composite scalar particle ($\Phi\sim\sum_a\varphi_a\varphi_a^\dagger\sim H^\dagger H$), as well as the 21 fundamental right-handed SM fermions $\xi$.

The pNGB's $\Pi$ are packaged in the field, 
\begin{equation}
\Sigma \equiv {\rm Exp} \left[ i \frac{\eta^\prime}{\sqrt{N_f f}} \right]
{\rm Exp} \left[ \sum_a 2i \frac{X^a \Pi^a}{f} \right] \Sigma_0
\end{equation}
where $X^a_{ij}$ are the $2N_f^2 - N_f -1$ broken generators of $\SU(2N_f)$.  The matrix $\Sigma_0=\langle\Sigma\rangle$ satisfies $\Sigma_0^\dagger\Sigma_0=\mathbbm{1}$ consistent with the $\SU(2N_f) \rightarrow \Sp(2N_f)$ chiral symmetry breaking pattern. A convenient choice of basis is:
\begin{equation*}
    \Sigma_0 = \begin{pmatrix}
        A & ~ & ~ \\
        ~ & A & ~ \\
        ~ & ~ & A
    \end{pmatrix}, \ \ \ A = \begin{pmatrix}
        0 & 1 & 0 & 0 \\
        -1 & 0 & 0 & 0 \\
        0 & 0 & 0 & 1 \\
        0 & 0 & -1 & 0
    \end{pmatrix} .
\end{equation*}
The vacuum during weak confinement spontaneously breaks
\begin{equation}
\SU(3)_c\times \U(1)_Y \rightarrow \SU(2)_c\times \U(1)_Q
\end{equation}
such that the QCD triplet multiplets split into a doublet plus a singlet. The $\SU(3)_c$ and $\U(1)_Y$ generators can be written in the $\psi$ basis as
\begin{eqnarray}
    L^a & = & \text{diag}(0,\lambda^a,0,\lambda^a,0,\lambda^a) \\
    Y & = & \text{diag}\left( -\frac{1}{2},\frac{1}{6},\frac{1}{6},\frac{1}{6},-\frac{1}{2},\frac{1}{6},\frac{1}{6},\frac{1}{6},-\frac{1}{2},\frac{1}{6},\frac{1}{6},\frac{1}{6} \right) \nonumber
\end{eqnarray}
where $\lambda^a$, $a=1,\ldots,8$, are the $3 \times 3$ Gell-Mann matrices. The generators of $\SU(2)_c$ are the usual Pauli matrices, while the generator of $\U(1)_Q$ is a combination of hypercharge and the $\lambda^8$ generator of $\SU(3)_c$:
\begin{equation}
\begin{split}
    Q & = \frac{1}{\sqrt{3}}L^8-Y \\
    & = \text{diag}\left(\frac{1}{2},-\frac{1}{2},0,0,\frac{1}{2},-\frac{1}{2},0,0,\frac{1}{2},-\frac{1}{2},0,0  \right) .
\end{split}
\end{equation} 

The weak confined vacuum preserves a single exact global $\U(1)_r$ symmetry defined in \Cref{eq:r-charge}, $r = -2Y +B -L$.
The pions have $r=0$, and the $r$-charges of the remaining particles (as well as their representations under $\SU(2)_c\times \U(1)_Q$) are summarized in \Cref{tab:particle_charges}.  An asymmetry in $r$ in the confined phase maps on to asymmetries in $B$, $L$, and/or $Y$ in the deconfined phase.

\begin{table}[t]
    \begin{tabular}{| c || c c | c | c |}
        \hline
        ~ &  $\SU(2)_c$ & $\U(1)_Q$ & $\U(1)_r$ & Constituents \\
         \hline
%         $\ell_i$ & $\mathbf{1}$ & $\frac{1}{2}$ & 0 & ~ \\
%         $q_{Si}$ & $\mathbf{1}$ & $-\frac{1}{2}$ & 0 & ~ \\
%         $q_{Di}$ & $\mathbf{2}$ & 0 & 0 & ~\\
%         $\varphi_1$ & $\mathbf{1}$ & $-\frac{1}{2}$ & $-1$ & ~\\
%         $\varphi_2$ & $\mathbf{1}$ & $\frac{1}{2}$ & $1$ & ~ \\
         $\Psi_1^0$ & $\mathbf{1}$ & 0 & $-1$ & $\ell\varphi_1$ \\
         $\Psi_1^{0c}$ & $\mathbf{1}$ & 0 & $+1$ & $q_{S}\varphi_2$\\ 
         $\Psi_1^+$ & $\mathbf{1}$ & $1$ & $+1$ & $\ell\varphi_2$ \\
         $\Psi_1^-$ & $\mathbf{1}$ & $-1$ & $-1$ & $q_S\varphi_1$ \\
         $\Psi_2^+$ & $\mathbf{2}$ & $1/2$ & $+1$ & $q_D\varphi_2$ \\
         $\Psi_2^-$ & $\mathbf{2}$ & $-1/2$ & $-1$ & $q_D\varphi_1$\\
         $\Phi$ & $\mathbf{1}$ & 0 & 0 & $\varphi_a\varphi_a^\dagger$ \\        
         \hline
         $E_i$ & $\mathbf{1}$ & $-1$ & $-1$ & ~\\
         $U_{Si}$ & $\mathbf{1}$ & $+1$ & $+1$ & ~\\
         $U_{Di}$ & $\mathbf{2}$ & $1/2$ & $+1$ & ~ \\
         $D_{Si}$ & $\mathbf{1}$ & 0 & $-1$ & ~ \\
         $D_{Di}$ & $\mathbf{2}$ & $-1/2$& $-1$ & ~ \\
         \hline
    \end{tabular}
    \caption{Table indicating the quantum numbers of the composite Dirac fermions ($\Psi$), composite scalar ($\Phi$), their fundamental constituents, and the quantum numbers of the $\SU(2)_L$ singlet fermions.}
    \label{tab:particle_charges}
\end{table}

The SM QCD and hypercharge interactions and Yukawa couplings explicitly break the $\SU(12)$ chiral symmetry, resulting in masses for most of the $\Pi$'s that are $m_\Pi \ll \Lambda_W$, and in subleading corrections to the ${\cal O}(\Lambda_W)$ masses of the $\Psi$'s and $\Phi$ composites.  The Yukawa interactions further mix the $\Psi$ Dirac fermions with the fundamental $\xi$ singlets.

\subsection{Interactions}

In the limit where the SM QCD, hypercharge, and Yukawa couplings are neglected, the EFT contains interactions among the composites dictated by the chiral symmetry \cite{Berger_2019}:
\begin{equation}
\begin{split}
\mathcal{L}_{\text{IR}} \supset & -\left( y_0\Lambda_W \right) \sum_{i,j=1}^{12}\sum_{a,b=1}^2\left[ \epsilon^{ab}\Psi_{ia}\Sigma_{ij}^\dagger\Psi_{jb} \right] +\text{h.c.} \\
& - \left(\frac{y'_0\Lambda_W}{f} \right) \sum_{i,j=1}^{12}\sum_{a,b=1}^2\left[ \epsilon^{ab}\Psi_{ia}\Sigma_{ij}^\dagger\Psi_{jb}\Phi\right] +\text{h.c.} 
\label{eq:WCSM-lagranian}
\end{split}
\end{equation}
where $y_0$ and $y'_0$ are $\mathcal{O}(1)$ complex dimensionless coefficients encapsulating the strong dynamics.  The $\Psi$ particles pair up into Dirac states with common mass $m_\Psi = y_0 \Lambda_W$.

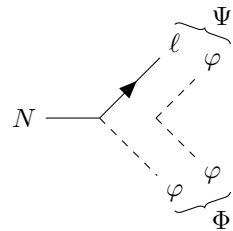
\begin{figure}[t]
\centering
%\begin{subfigure}[c]{0.2\textwidth}
\begin{tikzpicture}
\begin{feynman}
    \vertex at (0, 0) (i) {\(N\)};
    \vertex at (1,0) (a);
    \vertex at (2,1) (f1) {\(\ell\)};
    \vertex at (2,-1) (f2) {\(\varphi\)};
    \vertex at (1.75,0) (b);
    \vertex at (2.5,0.75) (f3) {$\varphi$};
    \vertex at (2.5,-0.75) (f4) {$\varphi$};
    \diagram*{
        (i) -- (a),
        (a) -- [fermion] (f1), 
        (a) -- [scalar] (f2),
        (b) -- [scalar] (f3),
        (b) -- [scalar] (f4),
    };
    \draw [decoration={brace}, decorate] (f1.north) -- (f3.north east) node [pos=0.5, above right] {$\Psi$};
    \draw [decoration={brace}, decorate] (f4.south east) -- (f2.south) node [pos=0.5, below right] {$\Phi$};
\end{feynman}
\end{tikzpicture}
%\caption{}
%\label{subfig:feyn-diag-SM-add-phi}
%\end{subfigure}
\caption{Schematic representation showing the underlying neutrino $N$-$\ell$-$\varphi$ Yukawa interaction dressed by a bystander $\SU(2)_L$ Higgs doublet during weak confinement, resulting in a $N$-$\Psi$-$\Phi$ interaction.}
\label{fig:motivating-feyn-diag}
\end{figure}

The Yukawa interactions of \cref{eq:Lseesaw} manifest in the confined phase in two significant ways.  First, they introduce mixing between the $\Psi^0_1$ fermions and the Majorana $N$'s, in analogy with the $\xi$-$\Psi$ mixing mentioned above.  Second, they result in couplings between $\Psi^0_1$, $N$, and $\Phi$:
\begin{equation}
\begin{split}
    & -C_\lambda \sum_{\alpha,\beta=1}^{3}\left[ \lambda_{\alpha\beta}^\nu N_{\alpha}\Psi_{1\beta}^0\Phi \right] +\text{h.c.} \\
\end{split}
\label{eq:WCSM-right-neutrino-weyl}
\end{equation}
where $C_\lambda$ is an $\mathcal{O}(1)$ complex parameter.  One can understand this term in the EFT heuristically as arising from the Yukawa interaction dressed up with a bystander SM Higgs, see \Cref{fig:motivating-feyn-diag}.

\section{Baryon Asymmetry}
\label{sec:cosmology}

During $\SU(2)$ confinement, the $N$ fields interact with the massive $\Psi^0_1$ and $\Phi$ fields.  We assume that the strong dynamics produces masses such that $m_\Phi \gtrsim m_{\Psi} \equiv y_0 \Lambda_W$, and thus it makes sense to consider $\Phi$ decaying into one flavor of $\Psi^0_1$ and one $N$ (or one $\overline{\Psi}^0_1$ and one $N$), through the Yukawa interaction, \cref{eq:WCSM-right-neutrino-weyl}.  At the loop level, $CP$-violating phases can bias the decay in favor of either $\Psi^0_1$ or $\overline{\Psi}^0_1$, resulting in production of a net $r$-charge in the $\Psi^0_1$ sector.

\subsection{Boltzmann Equations}
\label{sec:boltzmann-eq}

We construct a set of Boltzmann equations which encapsulate the dynamics leading to the evolution of the number densities of the relevant species in the plasma during weak confinement.  We work with the scaled number densities per comoving volume \cite{Kolb}:
\begin{equation}
    Y_i \equiv \frac{n_i}{s} ,
\end{equation}
where $n_i = \int d^3 p_i f$ is the particle number density of species $i$, and $s = \frac{2 \pi^2}{45} g_* T^3$ is the entropy density of the universe.  During weak confinement, the light degrees of freedom are composed of the pions, the $\SU(2)_C \times \U(1)_Q$ gauge bosons, and the $\SU(2)$ singlet fermions, leading to $g_* \simeq 110$.

The coupled Boltzmann equations governing the evolution of  $Y_\Phi$  and $Y_{\Delta r} = Y_{\Delta \Psi} = (Y_{\Psi} - Y_{\overline{\Psi}})$ take the form \cite{csfong2013}:
\begin{eqnarray}
    \frac{dY_\Phi}{dz} &=& -D_1(Y_\Phi - Y^{eq}_\Phi) , \\
    \frac{dY_{\Delta r}}{dz} &=& -\epsilon D_1(Y_\Phi - Y^{eq}_\Phi) - W_1 Y_{\Delta r} ,
\end{eqnarray}
where $z \equiv m_\Phi / T$, and $W_1$ and $D_1$ represent washout and decay terms respectively, given by
\begin{eqnarray}
    D_1(z) &=& \frac{\gamma_{\text{tot}} z}{s H(m_\Phi)} = K_1 z \frac{\mathcal{K}_1(z)}{\mathcal{K}_2(z)} , \\
    W_1(z) &=& \frac{1}{2} D_1(z)\frac{Y^{eq}_\Phi (z)}{Y^{eq}_r} ,
\end{eqnarray}
where $K_1$ is defined below in \cref{sec:washout}, $\mathcal{K}_{n}(z)$ are the $n$th order modified Bessel functions of the second kind, and $H(m_\Phi)$ is the Hubble scale evaluated at temperature $T = m_\Phi$.

The parameter $\epsilon$ tracks the asymmetry in the
$\Phi$ decays producing $\Psi$ and $\overline{\Psi}$.  For a particular flavor combination, this can be written:
\begin{equation}
    \label{eq:cpassym1}
    \epsilon_{\alpha\beta} \equiv \frac{\gamma(\Phi\rightarrow N_\alpha\Psi_\beta)-\gamma(\Phi\rightarrow N_\alpha\overline{\Psi}_\beta )}{\gamma_\text{tot}} .
\end{equation}
Summing over all flavors leads to a net asymmetry in the $\Phi$ decays of
$\epsilon \equiv \sum_{\alpha,\beta=1}^3\epsilon_{\alpha\beta}$.  

In \cref{eq:cpassym1}, $\gamma(i\rightarrow f)$ is the thermally averaged decay width, defined in the notation of Ref.~\cite{csfong2013} as
\begin{equation}
\begin{split}
    \gamma(i\rightarrow f) = &\int\frac{d^3\vec{p_i}}{(2\pi)^32E_i}\frac{d^3\vec{p_f}}{(2\pi)^32E_f}e^{E_i/T} \\
    & \times (2\pi)^4\delta^{(4)}(p_i-p_f) \left\vert \mathcal{A}(i\rightarrow f) \right\vert^2 \\
    % & = \frac{n_\Phi}{g*}\Gamma(\Phi) \frac{\mathcal{K}_2(z)}{\mathcal{K}_1(z)}
\end{split}
\end{equation}
where $\mathcal{A}(i\rightarrow f)$ is the decay amplitude, and $\gamma_{\text{tot}}$ is the thermally averaged decay width of $\Phi$ summed over all channels.

\subsection{Washout}
\label{sec:washout}

The total decay width of the $\Phi$ is the sum of decays into $\Psi \overline{\Psi}$ and $\Psi N$.  At leading order,
\begin{equation}
\begin{split}
    & \Gamma(\Phi \rightarrow \Psi_\alpha \overline{\Psi}_\alpha) = \frac{N_c}{4 \pi} \frac{(|g|^2 m^2_\Phi - (g+g^*)^2 m_\Psi^2)}{m_\Phi} \sqrt{1 - \frac{4 m_\Psi^2}{m_\Phi^2}} , \\
    & \Gamma(\Phi \rightarrow N_\alpha \Psi_\beta) = \frac{|Y_{\alpha\beta}|^2}{16 \pi} \frac{\left(m_\Phi^2 - m_\Psi^2\right)^2}{m_\Phi^3} , \\
    & \Gamma_\Phi = \sum_{\alpha \beta} \Gamma(\Phi \rightarrow N_\alpha \Psi_\beta) + \sum_{\alpha} \Gamma(\Phi \rightarrow \Psi_\alpha \overline{\Psi}_\alpha) ,
\end{split}
\end{equation}
where $g\equiv y'_0\Lambda_W/f$ is the $\Psi$-$\Psi$-$\Phi$ coupling for any $\Psi$, c.f. \cref{eq:WCSM-lagranian}, $N_c$ is the number of $\SU(2)_C$ colors of $\Psi$, and $Y_{\alpha\beta}\equiv C_\lambda\lambda_{\alpha\beta}$ is the strength of the $\Phi$-$N_\alpha$-$\Psi_{1\beta}^0$ interaction from \cref{eq:WCSM-right-neutrino-weyl}.  Note that the decays into $\Psi \overline{\Psi}$ include three flavors each of $\Psi^0_1 \Psi^{0c}_1$, $\Psi_1^+ \Psi_1^-$, as well as both color states of $\Psi_2^+ \Psi_2^-$ (12 channels total), and are only active provided $m_\Phi \geq m_\Psi /2$.

\begin{figure*}
    \centering    
    \includegraphics[width=\linewidth]{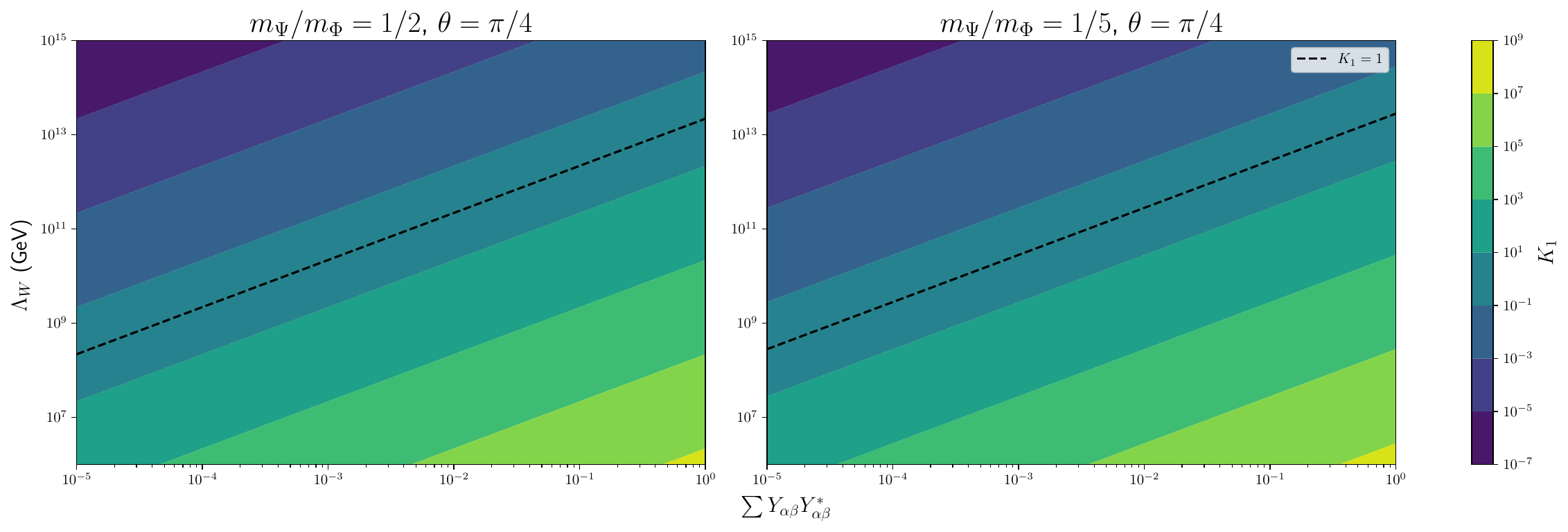}
    \caption{Washout strength $K_1$ in the plane of the Yukawa coupling $\sum |Y_{\alpha\beta}|^2$ and confinement scale $\Lambda_W$, for two values of $m_\Psi / m_\Phi$ and $\theta$, as indicated. Regions above the dotted line, $K_1 = 1$, are in the weak washout regime.}
    \label{fig:K1-plot}
\end{figure*}

Following the standard leptogenesis discussion, the strength of the washout is parameterized in terms of the quantity $K_1$:
\begin{equation}
    \begin{split}
    K_1  & = \frac{\gamma_{\text{tot}}}{s H(m_\Phi)} \frac{\mathcal{K}_2 (z)}{\mathcal{K}_1 (z)} \\
    & = \frac{1}{4 g^*} \left(\frac{1}{16 \pi} Y_{\alpha\beta}Y_{\alpha\beta}^* \frac{\left(m_\Phi^2 - m_\Psi^2\right)^2}{m_\Phi^3}\right)\left(\frac{3 M_{Pl}}{2 m_\Phi^2 \sqrt{\frac{g^*\pi^3}{5}}}\right) .        
    \end{split}
\end{equation}
Weak washout, for which the asymmetry generated in the $\Phi$ decays is expected to persist, is obtained for $K_1 \ll 1$.  In \Cref{fig:K1-plot} we plot $K_1$ in the plane of the complex Yukawa coupling $\sum |Y_{\alpha\beta}|^2$ and the confinement scale $\Lambda_W$, for two choices of the ratio $m_\Psi / m_\Phi$ and $\theta$ is the phase\footnote{In the basis where all of the other complex parameters have been chosen to be real.} of the coupling $g$. The dotted line indicates $K_1 = 1$, and weak washout is obtained for parameters above that line.

\subsection{CP Asymmetry}
\label{sec:cp-asym}

\begin{figure*}[t]
\begin{subfigure}{0.2\textwidth}
\begin{tikzpicture}
\begin{feynman}
    \vertex at (0, 0) (i) {\(\Phi\)};
    \vertex at (1,0) (a);
    \vertex at (2,1) (f1) {\(\Psi\)};
    \vertex at (2,-1) (f2) {\(N\)};
    \diagram*{
        (i) -- [scalar] (a),
        (a) -- [fermion] (f1), 
        (a) -- (f2),
    };
\end{feynman}
\end{tikzpicture}
\caption{}
\label{subfig:tree-level}
\end{subfigure}
\begin{subfigure}{0.2\textwidth}
\begin{tikzpicture}
\begin{feynman}
    \vertex at (0, 0) (i) {\(\Phi\)};
    \vertex at (1,0) (a);
    \vertex at (2,1) (b);
    \vertex at (2,-1) (c);
    \vertex at (3,1) (f1) {\(\Psi\)};
    \vertex at (3,-1) (f2) {\(N\)};
    \diagram*{
        (i) -- [scalar] (a),
        (a) -- [edge label=$N$] (b), 
        (a) -- [edge label=$\Psi$, fermion] (c),
        (b) -- [scalar, edge label=$\Phi$] (c),
        (b) -- [fermion] (f1),
        (c) -- (f2),
    };
\end{feynman}
\end{tikzpicture}
\caption{}
\label{subfig:psi-yukawa}
\end{subfigure}
\begin{subfigure}{0.2\textwidth}
\begin{tikzpicture}
\begin{feynman}
    \vertex at (0, 0) (i) {\(\Phi\)};
    \vertex at (1,0) (a);
    \vertex at (2,1) (b);
    \vertex at (2,-1) (c);
    \vertex at (3,1) (f1) {\(\Psi\)};
    \vertex at (3,-1) (f2) {\(N\)};
    \diagram*{
        (i) -- [scalar] (a),
        (a) -- [edge label=$N$] (b), 
        (c) -- [fermion, edge label=$\Psi$] (a),
        (b) -- [scalar, edge label=$\Phi$] (c),
        (b) -- [fermion] (f1),
        (c) -- (f2),
    };
\end{feynman}
\end{tikzpicture}
\caption{}
\label{subfig:antipsi-yukawa}
\end{subfigure}
\begin{subfigure}{0.2\textwidth}
\begin{tikzpicture}
\begin{feynman}
    \vertex at (0, 0) (i) {\(\Phi\)};
    \vertex at (1,0) (a);
    \vertex at (2,1) (b);
    \vertex at (2,-1) (c);
    \vertex at (3,1) (f1) {\(\Psi\)};
    \vertex at (3,-1) (f2) {\(N\)};
    \diagram*{
        (i) -- [scalar] (a),
        (a) -- [fermion, edge label=$\Psi$] (b), 
        (c) -- [fermion, edge label=$\Psi$] (a),
        (b) -- [scalar, edge label=$\Phi$] (c),
        (b) -- [fermion] (f1),
        (c) -- (f2),
    };
\end{feynman}
\end{tikzpicture}
\caption{}
\label{subfig:leading-loop}
\end{subfigure}
\caption{Feynman diagrams for the decay $\Phi \rightarrow \Psi N$, including the tree level diagram (a), higher order corrections induced by the Yukawa interactions [(b) and (c)], and from the $\Phi$-$\Psi$-$\Psi$ vertex (d).}
\label{fig:feynman-diagrams}
\end{figure*}
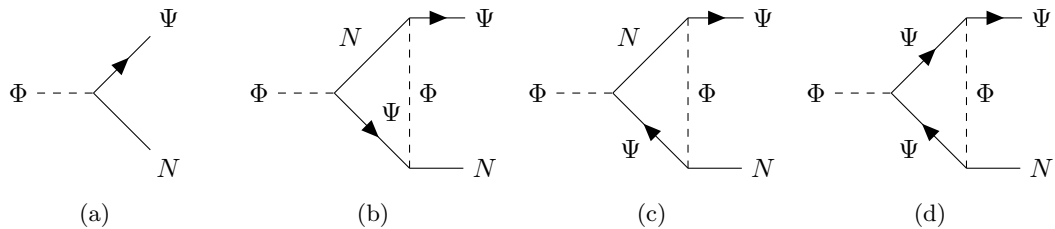

The flavor-dependent asymmetry $\epsilon_{\alpha \beta}$ vanishes at tree level, receiving its first non-zero contribution from the interference between the one-loop and tree level decay diagrams, shown in \cref{fig:feynman-diagrams}.  There are two types of one-loop corrections, proportional to $Y^3$ and $Y g^2$ (\cref{subfig:leading-loop}), respectively.  Since the expectation from naive dimensional analysis is that $g \sim \Lambda_W / f \gg Y$, we focus on the latter type here.

The thermal averaging factors are common to numerator and denominator, allowing us to rewrite \cref{eq:cpassym1} in terms of zero temperature quantities,
\begin{equation}
    \label{eq:cpassym2}
    \epsilon_{\alpha\beta} = \frac{\Gamma(\Phi\rightarrow N_\alpha\Psi_\beta)-\Gamma(\Phi\rightarrow N_\alpha\overline{\Psi}_\beta )}{\Gamma_\text{tot}} ,
\end{equation}
where to leading non-zero order the numerator should be evaluated at one loop, whereas the denominator can be approximated at tree level.  Summing over flavors, the total $CP$ asymmetry factor is
\begin{equation}
\begin{split}
    \epsilon
    & = \frac{1}{8\pi}\frac{k^2Y_{\alpha\beta}Y_{\alpha\beta}^*\Im[g^2](1-k^2)}{Y_{\alpha\beta}Y_{\alpha\beta}^*\frac{(1-k^2)^2}{\sqrt{1-4k^2}}+24(|g|^2-4k^2(\Re[g])^2)}
\end{split}
\end{equation}
where $k=\frac{m_\Psi}{m_{\Phi}}$ and we have assumed all of the $M_R$ are negligible compared to $m_\Phi$, as is motivated below. Notably, one can obtain a nonzero $\epsilon$ driven purely by a ``strong phase", $\Im[g^2]\neq 0$, which effectively decouples the $CP$ violation responsible for leptogenesis from $CP$-violation in the neutrino sector.  This requires $m_\Psi \leq m_\Phi / 2$, in order for the one-loop correction to realize an absorptive part.

\subsection{Evolution of the Plasma}

The $CP$ violation in the $\Phi$ decays produces an asymmetry between (one or more flavors of) $\Psi^0_1$ and its anti-particle, $\Psi^{0c}_1$, which can be characterized as an asymmetry in $r$-charge.  Rapid interactions between the $\Psi$'s and the pions will distribute the initial asymmetry in $\Psi^{0(c)}_1$ across the various flavors of $\Psi_1^\pm$ and $\Psi_2^\pm$ as well.  Eventually, the Universe cools sufficiently that the $\Psi$'s become non-relativistic and the heavier $\Psi_2^\pm$ and $\Psi_1^\pm$ decay back into the $\Psi_1^0$ states plus pions (which do not carry $r$-charge).  For simplicity, we assume that the period of $\SU(2)$ confinement persists long enough to reach this configuration (roughly $T \sim m_\Psi \sim k \Lambda_W$).

One way to understand the generation of the $\Delta r$ asymmetry is by assigning a putative $r$-charge of -1 to the $N$ fields.  The $\Phi$ decay into $N \Psi$ thus can be understood as storing $r$-charge in the $N$ sector, which is weakly coupled, and does not re-equilibrate.  This assignment is imprecise, because their Majorana masses imply that $N$ states can oscillate, effectively flipping the sign of the $r$-charge carried by $N$.  However, for $M_R \ll H(m_\Phi)$ (obtained for $\Lambda_W \gg M_R$), the oscillation time is Hubble-frozen until long after the $\Phi$'s have decayed, and one can engineer deconfinement to take place before $N$ oscillations can wash out the $\Delta r$ stored in the $\Psi$ particles.

\begin{figure}
\begin{tikzpicture}
\begin{feynman}
    \vertex at (0, 0) (i) {\(\Psi\)};
    \vertex at (1,0) (a);
    \vertex at (2,0.5) (c1) [crossed dot] {};
    \vertex at (3, 1) (f1) {\(\xi\)};
    \vertex at (2,-0.5) (b);
    \vertex at (3, -0.5) (c2) [crossed dot] {};
    \vertex at (4, -0.5) (f2)  {\(\xi\)};
    \vertex at (3, -1) (c3) [crossed dot] {};
    \vertex at (4, -1.5) (f3) {\(\overline{\xi}\)};
    \diagram*{
        (i) -- [fermion] (a),
        (a) -- [fermion, edge label = $\Psi$] (c1) -- [fermion] (f1), 
        (b) -- [scalar, edge label = $\Phi$] (a),
        (b) -- [fermion, edge label = $\Psi$] (c2) -- [fermion] (f2),
        (f3) -- [fermion] (c3) -- [fermion, edge label = $\overline{\Psi}$] (b) 
    };
\end{feynman}
\end{tikzpicture}
\caption{Feynman diagram indicating schematically how $\Psi$ decays into 3  singlets, $\xi$. The $\otimes$ symbol denotes an insertion of the mixing of a $\Psi$ into a $\xi$.}
\label{fig:xsi-feynman-diagrams}
\end{figure}
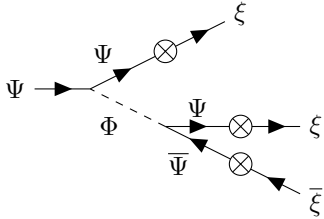

If weak confinement were to last long enough, the mixing between the $\Psi_1^0$ and the SM right-handed fermions (induced by the SM Yukawa interactions) could allow for decays into three right-handed quarks (shown in \Cref{fig:xsi-feynman-diagrams}), which are effectively massless during this epoch.  
The mixing angle between a $\Psi$ and one of the $\xi$ is \cite{Berger_2019}
\begin{equation}
    \theta_{f} \sim \frac{C_\lambda \lambda_{f}}{y_0 4\pi} \sim \frac{\lambda_{f}}{4 \pi} ,
\end{equation}
where $\lambda_{f}$ is the quark Yukawa interaction for the corresponding flavor of $\Psi$.  The 3-body decay is thus maximized for a bottom-flavored $\Psi^0_1$ decaying into a right-handed bottom quark and a pair of right-handed top quarks: 
\begin{equation}
    \Gamma_{\text{max}}(\Psi\rightarrow\xi\bar{\xi}\xi) = \frac{\theta_b^2 \theta_t^4}{192 (2\pi)^3}\left(\frac{m_\Psi}{m_\Phi}\right)^4 m_\Psi .
\end{equation}

\begin{figure}
    \centering    
    \includegraphics[width=0.99\linewidth]{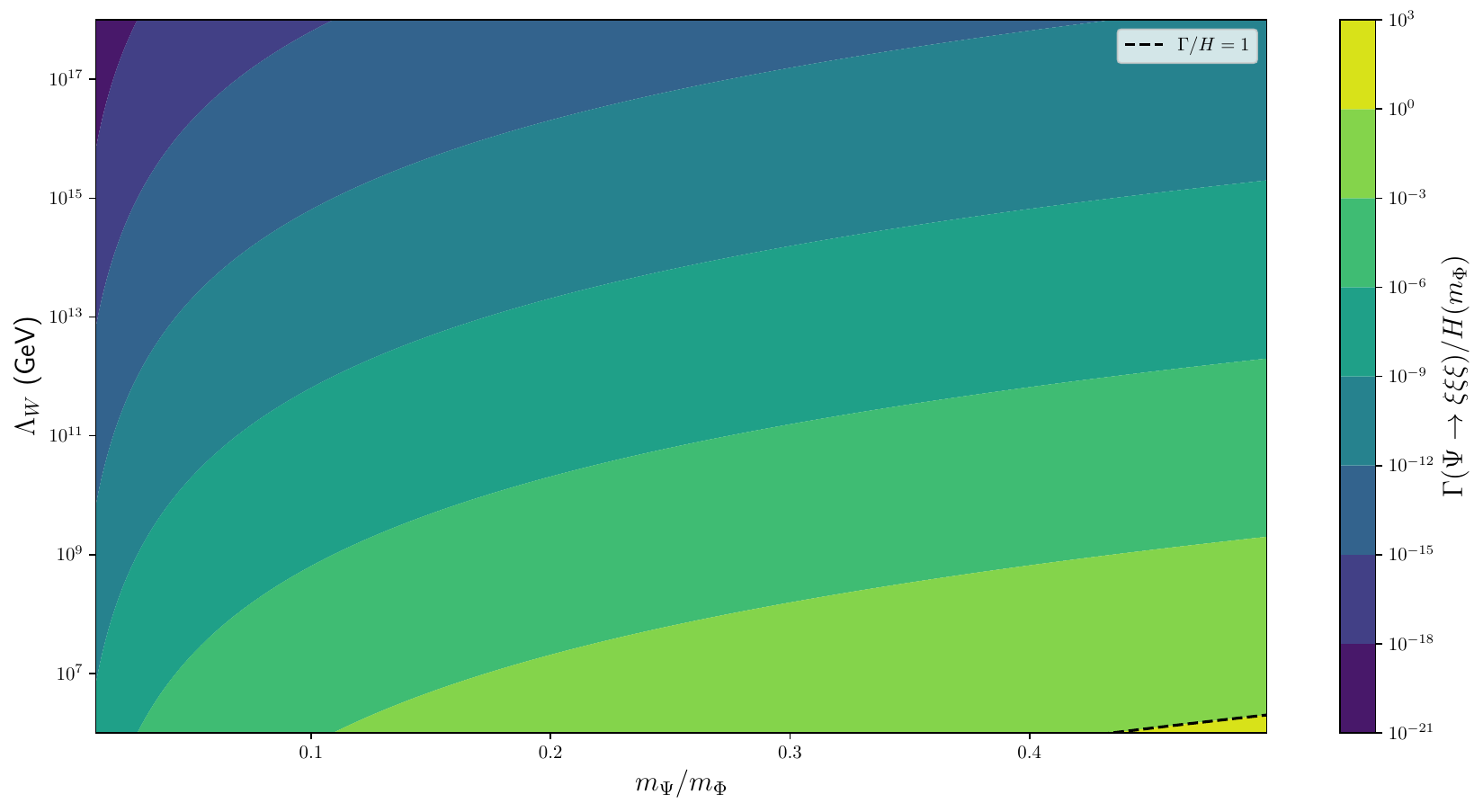}
    \caption{The ratio of the decay width $\Gamma(\Psi\rightarrow\xi\xi\xi)$} to the Hubble Scale $H(m_\Phi$) in the plane of $\Lambda_W$ and $m_\Psi/m_\Phi$. The dotted line indicates $\Gamma / H = 1$.)
    \label{fig:decaywidth}
\end{figure}

In \Cref{fig:decaywidth}, we plot the ratio of the resulting decay into singlet quarks divided by the Hubble scale.  We find that the temperatures at which the decay width is comparable to the Hubble scale are 7-11 orders of magnitude smaller than $m_\Phi \sim \Lambda_W$ itself, indicating that weak confinement would have to persist for a very long time for these decays to be relevant. Thus, we proceed by assuming that deconfinement occurs before these decays occur, and that all of the $r$-charge asymmetry will reside in the $\Psi^{0(c)}_1$ states at the time of deconfinement.
 
\subsection{Deconfinement} 
\label{sec:boltzmann-eq-basym}

Assuming $m_\Phi$ is slightly smaller than $\Lambda_W$, its number density at the time of decay will be very close to its equilibrium density in the confined plasma:
\begin{equation}
    Y^{\text{eq}}_{\Phi} = \frac{45}{2 \pi^4 g^*} z^2 \mathcal{K}_2(z) j_i \simeq \frac{90}{\pi^4 g^*} .
\end{equation}
The asymmetry in $\Delta r$ produced by its decay can thus be written
\begin{equation}
    Y_{\Delta r} \simeq \epsilon Y^{\text{eq}}_\Phi .
\end{equation}
As discussed above, for much of the parameter space this asymmetry will end up as an asymmetry in the lightest flavor of $\Psi_1^{0(c)}$.  Because of efficient annihilation into pions, for temperatures below $m_\Psi$, the asymmetric population of $\Psi_1^{0(c)}$ is likely to dominate over any tiny remnant symmetric population (though this detail does not impact the final baryon asymmetry).  This situation persists until the weak $\SU(2)_L$ coupling is restored to its observed zero-temperature value, resulting in deconfinement.

From \Cref{tab:particle_charges}, the valence parton content of $\Psi_1^0$ and $\Psi_1^{0c}$ is $\Psi_1^0 \sim \{\ell \varphi_1\}$, and $\Psi_1^{0c} \sim \{q_{S} \varphi_2\}$.  Thus, deconfinement will largely\footnote{Any asymmetry in hypercharge will quickly transfer into $B$ and/or $L$, since $\U(1)_Y$ is restored.} transfer the asymmetry in $\Delta r$ either entirely into an asymmetry in $\Delta B$ or $\Delta L$.  Either way, this represents an asymmetry in $B-L$, which cannot be erased by the weak sphalerons, which remain active until the electroweak phase transition.  

Since $M_R$ must be $\ll \Lambda_W$ in order to prevent washing out the $\Delta r$ asymmetry after the $\Phi$ decays, one concern could be that after deconfinement the $N$ decays themselves could contribute to the $B-L$ asymmetry, potentially washing it out.  However, the asymmetry in the $N$ decays arises purely in the Yukawa sector (see e.g. \cite{csfong2013} for a review), which is distinct from the strong phase in the coupling $g$ which we have invoked to generate the $\Delta r$ from $\Phi$ decays.  Thus, this scenario is much less reliant on the presence of cosmologically relevant $CP$-violation in the lepton sector than leptogenesis during a standard cosmological history. 

\begin{figure*}
    \centering
    \includegraphics[width=0.95\textwidth]{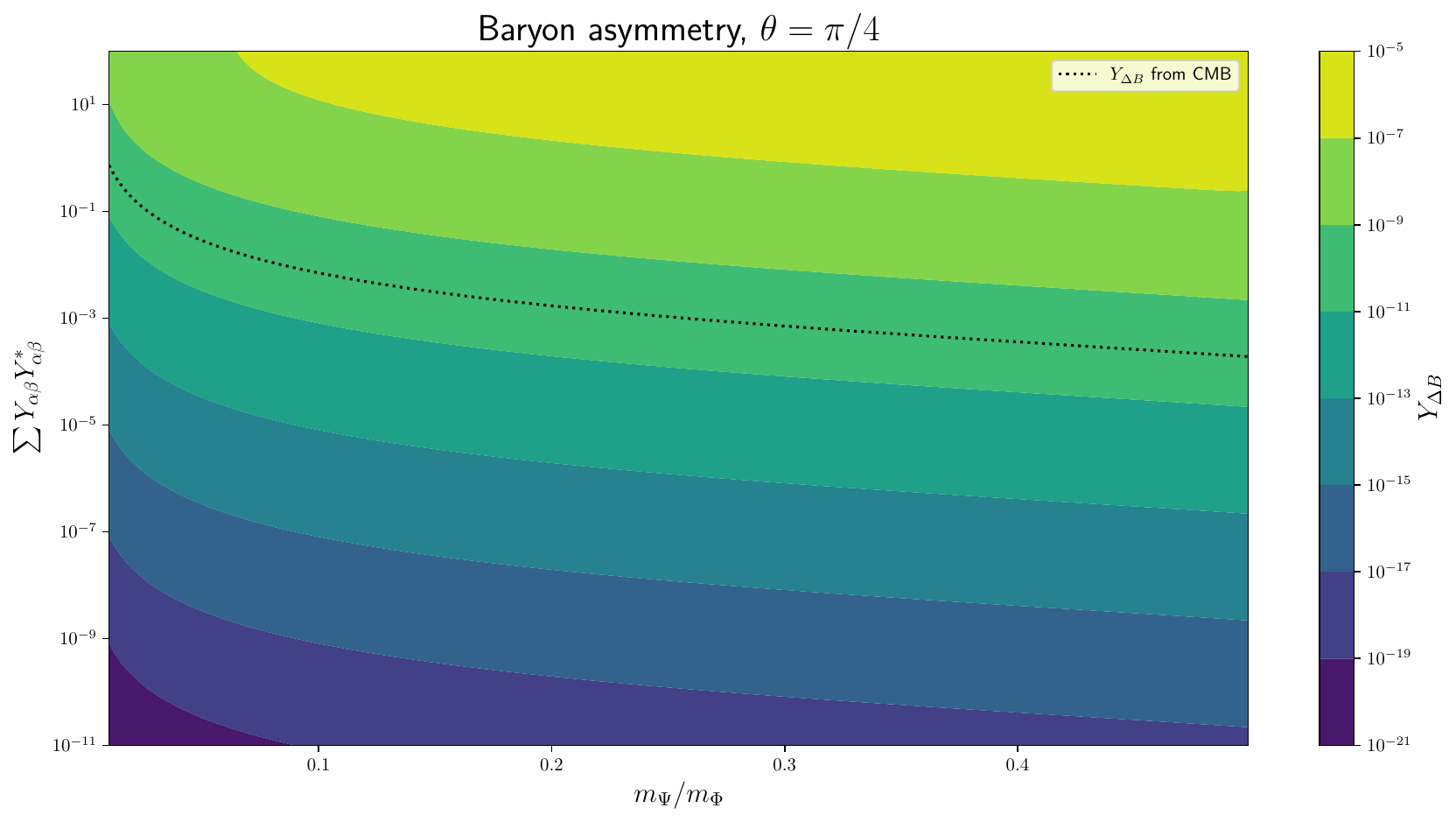}
    \caption{Resulting baryon asymmetry $Y_{\Delta B}$ in the plane of $m_\Psi / m_\Phi$ and $\sum |Y|^2$. The observed baryon asymmetry is indicated by the dotted line.}
    \label{fig:baryon-asymmetry}
\end{figure*}

Putting everything together, the final baryon asymmetry is given by the equilibrium conditions distributing $B-L$ in the SM plasma \cite{Turner1990}: 
\begin{equation}
\label{eq:baryon-asym}
    Y_{\Delta B}(\infty) = \frac{28}{79}Y_{\Delta (B-L)} = -\frac{28}{79}Y_{\Delta r} .
\end{equation}
In \Cref{fig:baryon-asymmetry}, we plot the resulting baryon asymmetry in the plane of the Yukawa couplings $\sum |Y|^2$ versus the ratio of masses, $m_\Psi / m_\Phi$, for $\theta = \pi/4$.  The dotted line indicates parameters matching the observed baryon asymmetry, \Cref{eq:YDB-experiment}.  Provided that $\Lambda_W$ is large enough so as to be in the weak washout regime (see \Cref{fig:K1-plot}), the asymmetry is not very sensitive to its precise value.  For Yukawa couplings resulting in the observed baryon asymmetry ($10^{-2}-1$, depending on $m_\Psi / m_\Phi$), this corresponds to $\Lambda_W \gtrsim 10^{11}$~GeV to $10^{14}$ GeV.  In the absence of fine-tuning, these Yukawas lead to light active neutrino masses similar to the observed mass differences for $M_R \sim 10^{10}$~GeV to $10^{14}$~GeV.

\section{Conclusion}
\label{sec:conclusion}

We examine a rather standard model of leptogenesis including a set of three singlet right-handed neutrinos with Majorana masses, but taking place in the backdrop of an unusual cosmology in which the coupling of the SM $\SU(2)_L$ interaction is promoted to a dynamical quantity, briefly becoming so strong as to result in confinement during a very early epoch before the electroweak phase transition.  As a result, the mechanism by which a baryon asymmetry can be generated is completely different, relying on decays of a heavy composite scalar state into right-handed neutrinos and composites containing quarks and leptons.  The $CP$-violation is provided by a strong phase in the EFT describing the interactions of the composites, rather than in the neutrino Yukawa couplings.  Indeed, in order to not wash out the asymmetry generated during this process, it may be favorable to have minimal $CP$-violation in the Yukawas relevant during the subsequent right-handed neutrino decays.

Our results depend on our assumptions concerning the parameters of the low energy EFT describing the interactions among the composite particles.  While the parameters we adopt are motivated by naive dimensional analysis, it would be far preferable to have more solid estimates of their sizes from e.g. lattice simulations.  Our work thus highlights the need for reliable non-perturbative calculations as inputs to understanding physics beyond the Standard Model.

More broadly, our work demonstrates the fact that the inference of model parameters based on cosmic observations can be subject to large dependence on the assumption of a standard cosmological history.  This dependence leads to haziness obscuring our understanding of where the most interesting regions of parameter space lie, which propagates into the high priority regions for experimental searches.  For example, measurements of $CP$-violation in neutrino oscillations are often held up as informing the viability of leptogenesis models.  Our work relies on a strong $CP$ phase from $\SU(2)_L$ confinement, and finds that in such a cosmology it may be advantageous to have negligible $CP$-violation in the right-handed neutrino decays.

\acknowledgements

TMPT is glad to acknowledge helpful conversations with Seyda Ipek and Jessica Turner.
The work of TMPT is supported in part by the U.S.\ National Science Foundation under Grant PHY-2210283.  This work was performed in part at Aspen Center for Physics, which is supported by National Science Foundation grant PHY-2210452 and the Kavli Institute for Theoretical Physics (KITP), supported in part by grant NSF PHY-2309135. 
The work of IG is supported in part by the National Science Foundation Graduate Research Fellowship Program (NSF GRFP) under grant DGE-1839285.

% Bibliography for the paper
\bibliography{bibliography}
\end{document}